\documentclass[
reprint,
superscriptaddress,
amsmath,amssymb,
aps,
pra,
]{revtex4-1}

\usepackage{graphicx}
\usepackage{dcolumn}
\usepackage{upgreek}
\usepackage{mathrsfs}
\usepackage{bm}
\usepackage[breaklinks=true,colorlinks=true,linkcolor=blue,urlcolor=blue,citecolor=blue]{hyperref}
\allowdisplaybreaks[4]
\begin{document}

\preprint{APS/123-QED}


\title{Two-Color Optical Nonlinearity in an Ultracold Rydberg Atom Gas Mixture}



\author{Cheng Chen}
\thanks{These authors contributed equally to this work}
\author{Fan Yang}
\thanks{These authors contributed equally to this work}
\author{Xiaoling Wu}
\author{Chuyang Shen}
\affiliation{State Key Laboratory of Low-Dimensional Quantum Physics, Department of Physics, Tsinghua University, Beijing 100084, China}
\author{Meng Khoon Tey}
\email{mengkhoon\_tey@mail.tsinghua.edu.cn}
\affiliation{State Key Laboratory of Low-Dimensional Quantum Physics, Department of Physics, Tsinghua University, Beijing 100084, China}
\affiliation{Frontier Science Center for Quantum Information, Beijing 100084, China}
\author{Li You}
\email{lyou@mail.tsinghua.edu.cn}
\affiliation{State Key Laboratory of Low-Dimensional Quantum Physics, Department of Physics, Tsinghua University, Beijing 100084, China}
\affiliation{Frontier Science Center for Quantum Information, Beijing 100084, China}
\affiliation{Beijing Academy of Quantum Information Sciences, Beijing 100193, China}



\begin{abstract}

We report the experimental observation of strong two-color optical nonlinearity in an ultracold gas of $^{85}\mathrm{Rb}$-$^{87}\mathrm{Rb}$ atom mixture. By simultaneously coupling two probe transitions of $^{85}$Rb and $^{87}$Rb atoms to Rydberg states in electromagnetically induced transparency (EIT) configurations, we observe significant suppression of the transparency resonance for one probe field when the second probe field is detuned at $\sim1~\mathrm{GHz}$ and hitting the EIT resonance of the other isotope. Such a cross-absorption modulation to the beam propagation dynamics can be described by two coupled nonlinear wave equations we develope. We further demonstrate that the two-color optical nonlinearity can be tuned by varying the density ratio of different atomic isotopes, which highlights its potential for exploring strongly interacting multi-component fluids of light.
\end{abstract}


\maketitle


\section{Introduction}\label{sec:sec1}
Cold atomic ensemble coupled to Rydberg states in the ladder configuration of electromagnetically induced transparency (EIT) \cite{Friedler2005} represents a promising platform for exploring quantum nonlinear optics \cite{Gorshkov2011,petrosyan2011electro,Bienias2016,Yang2019,Yang2020}, where giant nonlinear optical responses modifying light propagation can occur at the single-photon level \cite{saffman2010quantum,Peyronel2012,firstenberg2016non,murray2016quantum,wu2020concise}. Aimed at a full control of this effective photon-photon interaction, many experiments have been carried out leading to the realization of efficient single-photon emission \cite{Dudin2012,Ripka2018}, single-photon transistor \cite{Baur2014a,Tiarks2014,Gorniaczyk2014,Gorniaczyk2016}, photonic phase gate \cite{Tiarks2016,thompson2017symmetry,Tiarks2019}, and the creation of exotic few-photon quantum states \cite{Firstenberg2013,Liang2018,Cantu2020}, to name just a few.

At small optical depth per blockade radius, Rydberg EIT endows coherent photonic fields strong and nonlocal nonlinearities \cite{Ates2011,Sevincli2011Nonlocal,Stanojevic2013,Grankin2015,Bai2016}, as has been demonstrated by recent experiments ranging from the observed dissipative \cite{Pritchard2010,Boddeda2016,Han2016} and dispersive self-Kerr nonlinearity \cite{Parigi2012} to dispersive cross-Kerr nonlinearity between counter-propagating light beams \cite{Sinclair2019}. Most of the aforementioned experiments are, however, carried out in a single-species atomic ensemble, with optical nonlinearities limited to quantum probe field of one color, controlled by the upper leg strong classical coupling field. Although multi-color probe fields can be introduced with multi-color control fields, wave mixing from the consequent nonlinearities are complicated with no clear dominating pathways. In this work, we explore the atomic side by studying Rydberg EIT in a mixture gas ensemble of two atomic species. Our emphasis is  to establish interactions or correlations between spectrally distinguishable probe fields forming the EIT pathways respectively on the two atomic isotopes. Such a setup could benefit a variety of applications like all-optical switching in a multi-wavelength network \cite{stern2009multiwavelength} and support investigations of multi-component photon fluid dynamics \cite{Friedler2005pulse,situ2020dynamics}.

This Letter reports an unambiguous demonstration of dissipative cross-Kerr nonlinearity between probe fields of two different colors in a mixture gas ensemble of ultracold $^{85}$Rb and $^{87}$Rb atoms. By coupling light fields of two different frequencies respectively to Rydberg states of $^{85}$Rb and $^{87}$Rb atoms via two distinct EIT pathways, the Rydberg interaction between different species of atoms is mapped onto photonic interaction between the two-color optical probe fields. In the classical regime, this effective interaction manifests in a cross-absorption modulation: the on-resonance transmission of one probe field (for instance for $^{87}$Rb atoms) propagating through the gas mixture is strongly affected by the other probe field (for $^{85}$Rb atoms) as we observe, and vice versa. We extend the previously adopted one-color analytical model to two probe fields in the gas mixture, use it to describe the underlying nonlinear beam dynamics and find good agreement with experimental observations. We further verify that both self-Kerr and cross-Kerr nonlinearities in our system can be controlled by adjusting the density ratio of the two atomic species, and this leads to a promising implementation for an efficient all-optical switch featuring large extinction ratio and small insertion loss.

\begin{figure}
\centering
\includegraphics[width=\linewidth]{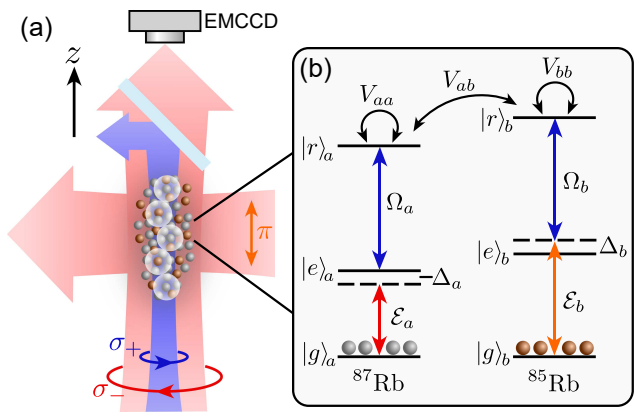}
\caption{(a) Schematic of the optical setup. A magnetic field ${B}=4.75~\mathrm{G}$ pointing upward is applied to specify the quantization axis $z$. (b) Schematics of the atomic level diagram. A 780-nm probe light $\mathcal{E}_a$ of $\sigma_{-}$ polarization couples state ${|g\rangle}_a = |5S_{1/2},F=2,m_F=-2\rangle$ to ${|e\rangle}_a = |5P_{3/2},F = 3,m_F = -3\rangle$ of $^{87}$Rb atoms, while a second probe light $\mathcal{E}_b$ of $\pi$ polarization connects $^{85}$Rb atoms from state ${|g\rangle}_b =|5S_{1/2},F=3,m_F=-3\rangle$ to ${|e\rangle}_b = |5P_{3/2},F=4,m_F=-3\rangle$. A 480-nm $\sigma_{+}$ polarized control field $\Omega$ drives both $^{87}$Rb and $^{85}$Rb atoms from intermediate states ${|e\rangle}_{a}$ and ${|e\rangle}_b$ to Rydberg states $|43S_{1/2},m_J=-1/2\rangle$ (${|r\rangle}_a$ and ${|r\rangle}_b$) with Rabi frequencies $\Omega_a$ and $\Omega_b=0.79\Omega_a$. The control field is kept on resonance with the $\left|e\right\rangle_a \leftrightarrow \left|r\right\rangle_a$ transition, while the probe field $\mathcal{E}_b$ is detuned by $\Delta_{b}/2\pi=4.0~\mathrm{MHz}$ from the ${|g\rangle}_b \leftrightarrow {|e\rangle}_b$ transition to satisfy the two-photon resonance condition for $^{85}$Rb. The beam waists for fields $\mathcal{E}_a$, $\mathcal{E}_b$, and $\Omega$ are $460~\mu\mathrm{m}$, $2030~\mu\mathrm{m}$, and $27~\mu\mathrm{m}$, respectively. The radial radius of the atomic ensemble is estimated to be $\sim100~\mu\mathrm{m}$.}\label{fig:fig1}
\end{figure}

\section{Experimental Setup}\label{sec:sec2}
Our experimental setup and the corresponding atomic level schemes are shown in Figs.~\ref{fig:fig1}(a) and \ref{fig:fig1}(b), respectively. To prepare the ultracold $^{85}$Rb-$^{87}$Rb mixture used in the experiment, both atomic isotopes captured by a magneto-optical trap (MOT) are first cooled with optical molasses and then prepared in their corresponding low-field seeking ground states $^{85}\text{Rb}\left|5S_{1/2},F=2,m_F=-2\right\rangle$ and $^{87}\text{Rb}\left|5S_{1/2},F=1,m_F=-1 \right\rangle$ via optical pumping. After that, the ultracold gas mixture is transferred into a magnetic quadrupole trap, followed by forced microwave evaporation of $^{87}$Rb atoms which sympathetically cools $^{85}$Rb atoms at the same time. The mixture is then loaded into an optical dipole trap (ODT) formed by a 1064-nm light beam and transferred to the desired initial states $^{85}\text{Rb}\left|5S_{1/2},F=3,m_F=-3\right\rangle$ and $^{87}\text{Rb}\left|5S_{1/2},F=2,m_F=-2\right\rangle$ via microwave driven adiabatic passages. Subsequently, the ultracold mixture is transported to a science chamber 21 cm away from the MOT chamber and loaded into a 1D optical lattice \cite{Shen2020}. At this stage, we have typically $2.2\times10^5$ $^{85}$Rb atoms and $7.1\times10^5$ $^{87}$Rb atoms at a temperature of $\sim$10 $\mu$K. The atomic cloud is then released for free expansion before the light fields forming EIT are shined on and transmission measurements are carried out. By changing the MOT loading time, ODT trap depth, and the free-expansion duration, we can vary atomic densities for different species. More details about the preparation of the ultracold mixture of $^{85}$Rb and $^{87}$Rb atoms have been described previously \cite{Dong2016,Cui2017}.

Once the ultracold mixture is prepared, we drive the two Rydberg EIT transitions with optical pathways shown in Fig.~\ref{fig:fig1}. Ground-state $^{85}$Rb and $^{87}$Rb atoms are respectively excited by a $\pi$ polarized probe light $\mathcal{E}_b$ and a $\sigma_-$ polarized probe light $\mathcal{E}_a$, which propagate perpendicularly to each other. Due to the difference in energy levels for $^{85}$Rb and $^{87}$Rb atoms, the probe fields $\mathcal{E}_b$ and $\mathcal{E}_a$ near the 780 nm D2 resonances differ in frequency by $\omega_b-\omega_a\sim 1~\mathrm{GHz}$. As a result, crosstalks between the two probe fields do not occur in the absence of strong Rydberg atom interactions \cite{Zeng2017}. The 480-nm $\sigma_+$ polarized light co-propagating with $\mathcal{E}_a$ serves as a strong control field $\Omega$, which couples intermediate states of both atomic species to their Rydberg states with respective Rabi frequencies $\Omega_a$ and $\Omega_b$ [see Fig.~\ref{fig:fig1}(b)]. After passing through the atomic ensemble, the control light is separated from the probe light by a dichroic mirror with only $\mathcal{E}_a$ left to be imaged onto an electron-multiplying charge-coupled device (EMCCD) camera via an optical imaging system. In each experimental run, probe fields $\mathcal{E}_a$ and $\mathcal{E}_b$ are turned on simultaneously for 10 $\mu$s, during which the EMCCD camera is exposed to image the transmitted probe field $\mathcal{E}_a$. The intensity transmission $T=I/I_0$ of the probe light $\mathcal{E}_a$, which we focus on here, is extracted by taking the ratio of the transmitted light intensity in the presence of the atomic ensemble ($I$) to that of without atomic cloud ($I_0$). For better quantitative comparison, only light intensity located near the central region ($24~\mu\mathrm{m}\times24~\mu\mathrm{m}$ square) of the control light is extracted from the image.

\section{Results and Discussions}\label{sec:sec3}
For the double EIT configuration shown in Fig.~\ref{fig:fig1}(b), two types of Rydberg interactions exist in the gas mixture: the intra-species van der Waals interaction $V_{\sigma\sigma}(\mathbf{r})$ between atoms in the same internal states ${|r\rangle}_\sigma$ ($\sigma=a,b$), and the inter-species one $V_{ab}(\mathbf{r})$ between different species atoms respectively in ${|r\rangle}_a$ and ${|r\rangle}_b$. While giant self-Kerr nonlinearities from intra-species interactions are already studied extensively and observed in a variety of settings \cite{Pritchard2010}, the inter-species interaction $V_{ab}$ supports a previously untouched cross nonlinear response between fields $\mathcal{E}_a$ and $\mathcal{E}_b$. To observe the optical nonlinearities, we first measure transmission spectra of the system by scanning the single-photon detuning $\Delta_a$ through the resonance of the $\left|g\right\rangle_a \leftrightarrow \left|e\right\rangle_a$ transition while keeping the control light $\Omega$ on resonance with the $\left|e\right\rangle_a \leftrightarrow \left|r\right\rangle_a$ transition.
\begin{figure*}
	\centering
	\includegraphics[width=0.9\textwidth]{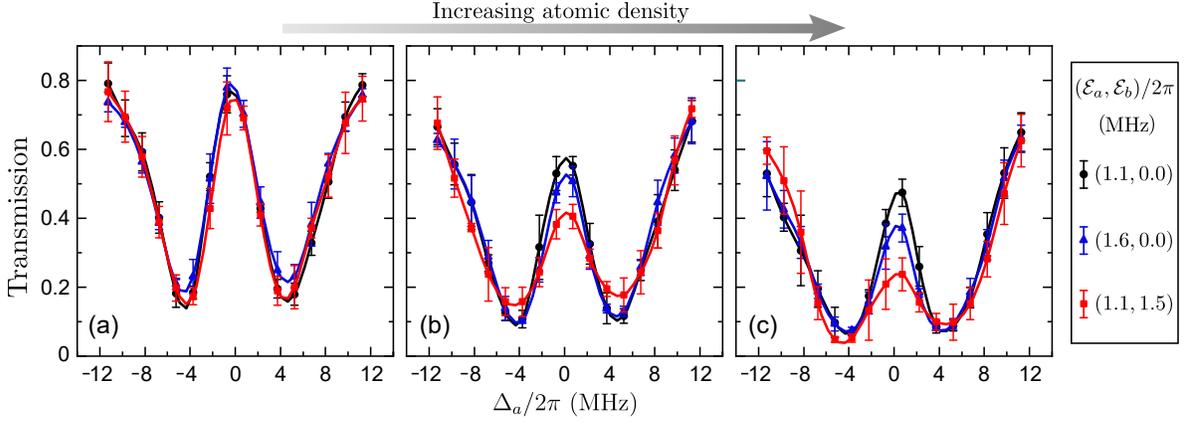}
	\caption{Transmission spectra of the probe light on $^{87}$Rb atoms at different atomic densities \cite{average}: (a) $\rho_{a}=4.1\times10^{10}$ cm$^{-3}$ ($^{87}$Rb) and $\rho_{b}=6.1\times10^{9}$ cm$^{-3}$ ($^{85}$Rb); (b) $\rho_{a}=4.8\times10^{10}$ cm$^{-3}$ and $\rho_{b}=1.2\times10^{10}$ cm$^{-3}$; (c) $\rho_{a}=5.9\times10^{10}$ cm$^{-3}$ and $\rho_{b}=2.0\times10^{10}$ cm$^{-3}$. The control field is kept at $\Omega_a/2\pi=$ 9.5 MHz, and Rabi frequencies for the input probe fields are $\mathcal{E}_a/2\pi=$ 1.1 MHz and $\mathcal{E}_b/2\pi=$ 0.0 MHz (black circles), $\mathcal{E}_a/2\pi=$ 1.6 MHz and $\mathcal{E}_b/2\pi=$ 0.0 MHz (blue triangles), $\mathcal{E}_a/2\pi=$ 1.1 MHz and $\mathcal{E}_b/2\pi=$1.5~MHz (red squares). The eye-guiding solid lines in (a)-(c) connecting the experimental data points come from spline fitting with the error bars denoting standard deviation from five measurements.}\label{fig:fig2}
\end{figure*}

The evolution of the transmission spectra with increasing atomic densities is shown in Fig.~\ref{fig:fig2} for three sets of probe field Rabi frequencies: (i) weak probe field $\mathcal{E}_a$ with vanishing probe field $\mathcal{E}_b$ (black circles); (ii) strong probe field $\mathcal{E}_a$ with vanishing probe field $\mathcal{E}_b$ (blue triangles); and (iii) weak probe field $\mathcal{E}_a$ with finite probe field $\mathcal{E}_b$ (red squares). At small atomic densities [see Fig.~\ref{fig:fig2}(a)], the observed EIT spectra for different input probe intensities (i)-(iii) almost completely overlap, indicating that the system is essentially linear. With increasing atomic densities, Rydberg interaction becomes prominent, which effectively shifts the Rydberg level and turns the on-resonance EIT to be off-resonant, i.e., changing the transparent ensemble into a two-level medium with large absorption. Consequenctly, the otherwise overlapped EIT spectra shifts away from each other and the EIT feature becomes asymmetric. As shown in Fig.~\ref{fig:fig2}(b), suppression of the on-resonance transmission ($\Delta_a=0$) induced by enhanced input intensity of $\mathcal{E}_a$ [comparing (i) and (ii)] arises from the dissipative self-nonlinearity generated by intra-species interaction $V_{aa}$, while the suppression induced by the other probe light $\mathcal{E}_b$ [comparing (i) and (iii)] reveals the existence of the cross optical nonlinearity between probe fields $\mathcal{E}_a$ and $\mathcal{E}_b$ generated by inter-species interaction $V_{ab}$. When atomic densities are further increased [see Fig.~\ref{fig:fig2}(c)], suppression of the transmission peak becomes more significant, owing to enhanced nonlinear effect.

To gain a quantitative understanding of the observed self- and cross-absorption modulations, we develop a theoretical model based on the cluster expansion method \cite{Sevincli2011Nonlocal,Tebben2019}. Retaining the nonlinearities to the third order, propagation of the probe fields $\mathcal{E}_a(\mathbf{r})$ and $\mathcal{E}_b(\mathbf{r})$ are found to be governed by a pair of coupled nonlinear wave equations
\begin{align}
i\partial_{\xi_a}
\mathcal{E}_a(\mathbf{r})
=&\left(-\frac{1}{2k_a}\nabla_{\perp_a}^2-\chi_a^{(1)}-\chi_a^{(\mathrm{3})}\right)\mathcal{E}_a(\mathbf{r}),\label{eq:eq1}\\
i\partial_{\xi_b}
\mathcal{E}_b(\mathbf{r})
=&\left(-\frac{1}{2k_b}\nabla_{\perp_b}^2-\chi_b^{(1)}-\chi_b^{(\mathrm{3})}\right)\mathcal{E}_b(\mathbf{r}),\label{eq:eq2}
\end{align}
where $\xi_{a(b)}$ denotes the direction of the wave vector $k_{a(b)}$ for the probe field $\mathcal{E}_{a(b)}$, and transverse gradient operator $\nabla_{\perp_{a(b)}}$ describes its diffraction perpendicular to $\xi_{a(b)}$. The linear susceptibility is denoted by $\chi^{(1)}_{a(b)}$, and the nonlinear ones are given by
\begin{align}
\chi_a^{(\mathrm{3})}=&\int d\mathbf{r}^\prime[\chi_{aa}^{(3)}(\mathbf{r}-\mathbf{r}^\prime)|\mathcal{E}_a(\mathbf{r}^\prime)|^2+\chi_{ab}^{(3)}(\mathbf{r}-\mathbf{r}^\prime)|\mathcal{E}_b(\mathbf{r}^\prime)|^2],\nonumber\\
\chi_b^{(\mathrm{3})}=&\int d\mathbf{r}^\prime[\chi_{ba}^{(3)}(\mathbf{r}-\mathbf{r}^\prime)|\mathcal{E}_a(\mathbf{r}^\prime)|^2+\chi_{bb}^{(3)}(\mathbf{r}-\mathbf{r}^\prime)|\mathcal{E}_b(\mathbf{r}^\prime)|^2].\nonumber
\end{align}
In the limit of negligible Rydberg excitation decoherence rate $\gamma$ and balanced coupling $\Omega_b\approx\Omega_a=\Omega$, the nonlocal third-order susceptibilities take the form
\begin{align}
\chi_{aa}^{(3)} &= \frac{i\sigma_0\Gamma_0\Gamma_aV_{aa}(\mathbf{r}-\mathbf{r}^\prime)/8\Omega^2}{(\Gamma_a^2+\Omega^2)V_{aa}(\mathbf{r}-\mathbf{r}^\prime)-2i\Gamma_a\Omega^2}\cdot\rho_a^2, \label{eq:eq3}\\
\chi_{ab}^{(3)} &= \frac{i\sigma_0\Gamma_0(\Gamma_a+\Gamma_b)V_{ab}(\mathbf{r}-\mathbf{r}^\prime)/{16\Omega^2}}{(\Gamma_a\Gamma_b+\Omega^2)V_{ab}(\mathbf{r}-\mathbf{r}^\prime)-i(\Gamma_a+\Gamma_b)\Omega^2}\cdot \rho_a\rho_b,\label{eq:eq4}
\end{align}
where $\sigma_0 = 3\lambda^2/2\pi$ and $\Gamma_0$ respectively denote the absorption cross section and the natural linewidth of the $^{87}$Rb D2 transition, $\Gamma_{a(b)}=\Gamma_{a({b})}^\prime-i\Delta_{a(b)}$ is the effective detuning with $\Gamma_{a({b})}^\prime$ the coherence decay rate of the probe transition containing the contributions from atomic spontaneous emission rate of the excited state and the linewidth of the probe laser, and $\rho_{a(b)}$ is the density of $^{87}$Rb ($^{85}$Rb) atoms.

\begin{figure}
	\centering
	\includegraphics[width=\linewidth]{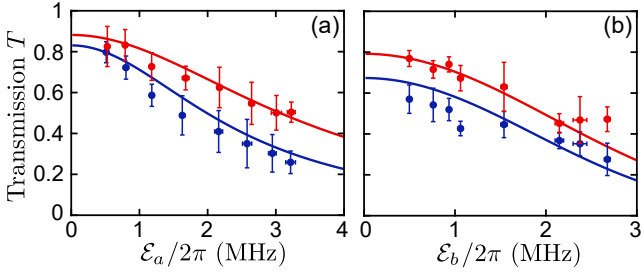}
	\caption{On-resonance ($\Delta_a=0$) transmission as a function of input Rabi frequencies $\mathcal{E}_a$ (a) and $\mathcal{E}_b$ (b) for $\mathcal{E}_b=0$ in (a) and $\mathcal{E}_a/2\pi=1.2~\mathrm{MHz}$ in (b). The blue (lower) and red (upper) dots are measured for different atomic densities $\rho_{a}=6.0\times10^{10}$ cm$^{-3}$, $\rho_{b}=3.2\times10^{10}$ cm$^{-3}$, $L=280~\mu\mathrm{m}$ (blue dots), and $\rho_{a}=4.2\times10^{10}$ cm$^{-3}$, $\rho_{b}=3.6\times10^{10}$ cm$^{-3}$, $L=273~\mu\mathrm{m}$ (red dots). The solid lines show  theoretical model predictions with $\Omega_a/2\pi=9.5~\mathrm{MHz}$, $\Gamma_{a}^\prime/2\pi=\Gamma_{b}^\prime/2\pi=3.6~\mathrm{MHz}$, and $\gamma/2\pi=300~\mathrm{kHz}$, while the error bars refer to standard deviation from seven measurements.}\label{fig:fig3}
\end{figure}
In our system (see Fig.~\ref{fig:fig1}), the control light $\Omega$ is tightly focused in the $x,y$ direction, hence propagation effect of the probe field $\mathcal{E}_b(\mathbf{r})$ is insignificant. Furthermore, in the near-resonant regime $|\Delta_{a(b)}|\ll\Gamma^\prime_{a(b)}$, nonlinear absorption dominates the beam dynamics, where both diffraction and nonlocal response can be neglected \cite{Sevincli2011Nonlocal}. With these approximations, we finally arrive at a simple propagation equation for the probe field $\mathcal{E}_a(\mathbf{r})$
\begin{equation}
\partial_{z}
\mathcal{E}_a(\mathbf{r})
=i\left[\chi_a^{(1)}+\chi_{aa}^{(\mathrm{eff})}|\mathcal{E}_a(\mathbf{r})|^2+\chi_{ab}^{(\mathrm{eff})}|\mathcal{E}_b(\mathbf{r})|^2\right]\mathcal{E}_a(\mathbf{r}),\label{eq:eq5}
\end{equation}
where the effective Kerr coefficients $\chi^{(\mathrm{eff})}_{aa}=\int d\mathbf{r}^\prime\chi^{(3)}_{aa}(\mathbf{r}-\mathbf{r}^\prime)$ and $\chi^{(\mathrm{eff})}_{ab}=\int d\mathbf{r}^\prime\chi^{(3)}_{ab}(\mathbf{r}-\mathbf{r}^\prime)$ are obtained by integrating over the original nonlocal susceptibilities. For probe field $\mathcal{E}_{a(b)}$ with a uniform transverse distribution, the solution to Eq.~(\ref{eq:eq5}) yields an analytical expression for the intensity transmission of $\mathcal{E}_{a}$
\begin{equation}
T=\frac{(\kappa_a+\kappa_{ab}I_b)e^{-(\kappa_a+\kappa_{ab}I_b)L}}{\kappa_a+\kappa_{ab}I_b+\kappa_{aa}I_a\left(1-e^{-(\kappa_a+\kappa_{ab}I_b)L}\right)} ,\label{eq:eq6}
\end{equation}
with $\kappa_a=2\mathrm{Im}[\chi_a^{(1)}]$, $\kappa_{aa}=2\mathrm{Im}[\chi_{aa}^{(\mathrm{eff})}]$, $\kappa_{ab}=2\mathrm{Im}[\chi_{ab}^{(\mathrm{eff})}]$, and $L$ the longitudinal length of the atomic cloud. The dependence of the transmission $T$ on the input probe field intensities $I_a=|\mathcal{E}_a|^2$ and $I_b=|\mathcal{E}_b|^2$ arise respectively from the self-Kerr and cross-Kerr nonlinearities, which are physical origins for the observed nonlinear absorption phenomena.

To verify the above theoretical model, we measure the on-resonance ($\Delta_a=0$) transmission as a function of input probe fields $\mathcal{E}_a$ and $\mathcal{E}_b$. For weak Rabi frequencies $\mathcal{E}_a,\mathcal{E}_b\ll\Omega$, our model shows good agreement with experimental results (see Fig.~\ref{fig:fig3}). In general, both ground-state atom densities and Rydberg excitation densities contribute to the nonlinearities, which lead to distinct dependence of self- and cross-nonlinear susceptibilities on $\rho_a$ and $\rho_b$, i.e., $\chi_{aa}^{(3)}\propto\rho_a^2$ and $\chi_{ab}^{(3)}\propto\rho_a\rho_b$ [see Eqs.~(\ref{eq:eq3}) and (\ref{eq:eq4})]. This suggests that the relative strength of the nonlinear coefficient $\chi_{aa}^{(3)}/\chi_{ab}^{(3)}$ can be tuned by varying the density ratio of the two atomic species. As confirmed in Figs.~\ref{fig:fig3}(a) and \ref{fig:fig3}(b), by decreasing $\rho_a$ while keeping $\rho_b$ fixed, a notable decrease of the slope in the $T$-$\mathcal{E}_a$ curve is observed, while the slope in the $T$-$\mathcal{E}_b$ curve experiences a much smaller change.
\begin{figure}
	\centering
	\includegraphics[width=\linewidth]{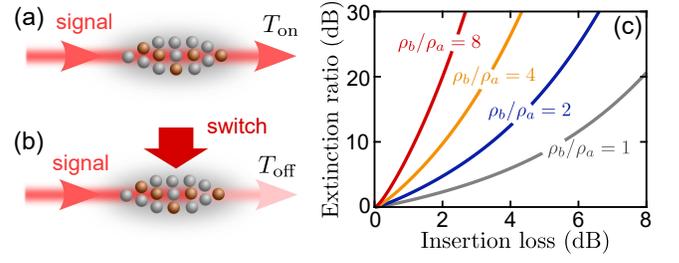}
	\caption{(a) and (b) show schematics of the on- and off-mode of an all-optical switch. (c) The extinction ratio for the proposed optical switch as a function of insertion loss (determined solely by $\rho_a$ here). For demonstration purpose, we assume that the phase noise of two probe fields can be neglected [i.e., $\Gamma_{a}^\prime/2\pi=\Gamma_{b}^\prime/2\pi=3~\mathrm{MHz}$] and Rydberg excitation has a small decoherence rate $\gamma/2\pi=100~\mathrm{kHz}$. The calculations are performed with $L=400~\mu\mathrm{m}$, $\Omega_a/2\pi=5~\mathrm{MHz}$, $\Delta_a=0$, $\Delta_b/2\pi=4~\mathrm{MHz}$, $\mathcal{E}_a/2\pi=1~\mathrm{MHz}$, and $\mathcal{E}_b/2\pi=1~\mathrm{MHz}$ [for the off-mode shown in (b)].}\label{fig:fig4}
\end{figure}

The tunablility of the self- and cross-nonlinearity based on manipulation of the density ratio demonstrated above can be applied to a variety of settings. Here, as a concrete example, we theoretically demonstrate that a flexible control of $\rho_b/\rho_a$ can be harnessed to improve the performance of a Rydberg-EIT based all-optical switch, where transmitted state of the signal light ($\mathcal{E}_a$) is controlled by the switch light field ($\mathcal{E}_b$) as depicted respectively in Figs.~\ref{fig:fig4}(a) and \ref{fig:fig4}(b) for the on- and off-mode of the switch. To realize such an optical switch, the probe field $\mathcal{E}_a$ needs to be tightly focused within the control field and the atomic ensemble. An efficient optical switch should possess large extinction ratio $10\log(T_\mathrm{on}/T_\mathrm{off})$ together with small insertion loss $10\log(1/T_\mathrm{on})$. Although such a switch can also be implemented in a single-species Rydberg EIT medium with $\mathcal{E}_a$ and $\mathcal{E}_b$ of the same frequency, a high extinction ratio from increasing atomic density in this case, is always accompanied by inevitable large insertion loss caused by enhanced self-nonlinearity. In our dual-species gas ensemble, such a trade-off can be significantly mitigated by generating unbalanced self-nonlinearity $\chi_{aa}^{(3)}$ and cross-nonlinearity $\chi_{ab}^{(3)}$ with unbalanced densities $\rho_a$ and $\rho_b$ as demonstrated. The switch performance for different density ratios are compared in Fig.~\ref{fig:fig4}(c). For balanced densities $\rho_a=\rho_b$ (such as the case of single-species), large extinction ratio likewise always implies severe insertion loss. By increasing the density ratio $\rho_b/\rho_a$ \cite{ratio}, improvements can be made, where high extinction ratio can be achieved at relatively low insertion loss. In the case of single-species, similar improvement occurs by choosing energetically far separated Rydberg levels for the signal and the switch field respectively \cite{Baur2014a,Tiarks2014,Gorniaczyk2014,Gorniaczyk2016}, where an extra control field is required.

\section{Conclusion and Outlook}\label{sec:sec4}
In conclusion, we have measured Rydberg EIT induced cross-Kerr nonlinearity between two nondegenerate optical fields in an ultracold $^{85}$Rb-$^{87}$Rb atomic mixture gas. The analytical model we develope provides an intuitive understanding of the nonlinear beam dynamics and is in quantitative agreement with experiments. We have demonstrated the tunability of the self-Kerr and cross-Kerr nonlinearities by varying the density ratio of different atomic species, and we have discussed their prospects for implementing a quality all-optical switch. This work highlights the ability of mixture Rydberg atom gas ensemble for exploring quantum nonlinear optics. Extending current studies to large intermediate state detunings, one can establish a system with efficient cross-phase modulation operating in the dispersive regime. By using pulsed or tightly focused input fields, it is also possible to realize temporal or spatial dark-bright soliton pairs \cite{park2000}, where the tunable nonlinearity allows one to address the interesting non-Manakov region \cite{chai2020magnetic}. Moreover, bringing the system to the single-photon regime, e.g., by using a higher-lying Rydberg state, could open up several new avenues, such as the construction of a single-photon transistor and the study of multi-component photonic Luttinger liquid \cite{Otterbach2011wigner}.

\begin{acknowledgments}
{\it Acknowledgments}.---We thank K. M{\o}lmer, T. Pohl, X. Liang, X. Chai, D. Lao, and Q. Zhang for valuable discussions. This work is supported by the National Key R$\&$D Program of China (Grants No.~2018YFA0306504 and No.~2018YFA0306503) and the National Natural Science Foundation of China (NSFC) (Grants No.~11654001, No.~91836302, and No.~U1930201). 

\end{acknowledgments}


\bibliography{ref}

\begin{thebibliography}{46}
\bibitem [{ave()}]{average}%
  \BibitemOpen
  \href@noop {} {}\bibinfo {note} {$\rho_a$ and $\rho_b$ are average densities
  along the central axis of the ensemble for $^{87}$Rb and $^{85}$Rb atoms,
  respectively.}\BibitemShut {Stop}%
\bibitem [{rat()}]{ratio}%
  \BibitemOpen
  \href@noop {} {}\bibinfo {note} {The density ratio cannot be arbitrarily
  large and is ultimately limited by $\rho_b$, which needs to be sufficiently
  small to validate the non-depletion approximation used to obtain
  Eq.~(\ref{eq:eq6})}\BibitemShut {NoStop}%
\end{thebibliography}

\end{document}